\pdfoutput=1
\documentclass[a4paper,fleqn,usenatbib]{mnras}

\usepackage[T1]{fontenc}
\usepackage{ae,aecompl}
\usepackage{graphicx}
\usepackage{physics}
\usepackage{amsmath}

\usepackage{graphicx}	
\usepackage{amsmath}	
\usepackage{amssymb}	
\newcommand{\msun}{{\rm M}_\odot}
\newcommand{\zsun}{{\rm Z}_\odot}
\usepackage{newtxtext,newtxmath}

\usepackage[normalem]{ulem}

\title[Evolution and explosions of metal enriched SMSs.]{Evolution and explosions of metal enriched supermassive stars: proton rich general relativistic instability supernovae. }

\author[C. Nagele et al.]{
Chris Nagele,$^{1}$\thanks{E-mail: chrisnagele.astro@gmail.com}
Hideyuki Umeda,$^{1}$
Koh Takahashi $^{2}$
\\
$^{1}$Department of Astronomy, Graduate School of Science, the University of Tokyo, Tokyo, 113-0033, Japan\\
$^{2}$Astronomical Institute, Graduate School of Science, Tohoku University, Sendai, 980-8578, Japan\\
}

\date{Accepted XXX. Received YYY; in original form ZZZ}

\pubyear{2022}

\begin{document}
\label{firstpage}
\pagerange{\pageref{firstpage}--\pageref{lastpage}}
\maketitle

\begin{abstract}
The assembly of supermassive black holes poses a challenge primarily because of observed quasars at high redshift, but additionally because of the current lack of observations of intermediate mass black holes. One plausible scenario for creating supermassive black holes is direct collapse triggered by the merger of two gas rich galaxies. This scenario allows the creation of supermassive stars with solar metallicity. We investigate the behavior of metal enriched supermassive stars which collapse due to the general relativistic radial instability during hydrogen burning. These stars contain both hydrogen and metals and thus may explode due to the CNO cycle (carbon-nitrogen-oxygen) and the rp process (rapid proton capture). We perform a suite of stellar evolution simulations for a range of masses and metallicities, with and without mass loss. We evaluate the stability of these supermassive stars by solving the pulsation equation in general relativity. When the stars becomes unstable, we perform 1D general relativistic hydrodynamical simulations coupled to a 153 isotope nuclear network with cooling from neutrino reactions, in order to determine if the stars explode. If the stars do explode, we post process the nucleosynthesis using a 514 isotope network which includes additional proton rich isotopes. These explosions are characterized by enhanced nitrogen and intermediate mass elements ($16\geq\rm{A}\geq25$), and suppressed light elements ($8\geq\rm{A}\geq14$), and we comment on recent observations of super-solar nitrogen in GN-z11.

\end{abstract}



\suppressfloats

\section{Introduction}
\label{introduction}

The study of supermassive stars has arisen from two peculiarities of the black hole population in the Universe. The first is that supermassive black holes (SMBHs) exist soon after the big bang \citep{mortlock2011,wu2015,banados2018,matsuoka2019,wang2021,Eilers2022arXiv221116261E,Marshall2023arXiv230204795M,Fan2022arXiv221206907F}. Our current understanding of cosmology requires that the SMBHs did not exist at the time of the big bang, implying that they were created in the brief intervening period. The second peculiarity is that the black hole mass function seems to be bimodal, with a noticeable lack of intermediate mass black holes (having masses in between solar mass black holes and SMBHs), although this may be due to observational bias \citep[][]{Wrobel2016AJ....152...22W,Baumgardt2017MNRAS.464.2174B,Kiziltan2017Natur.542..203K}. If the bimodality is not due to observational bias, it stands in opposition to the distributions of other self gravitating objects (stars and galaxies), which have continuous mass functions. 

The direct collapse black hole (DCBH) scenario was proposed to resolve the first of these peculiarities \citep{bromm2003}, and is sometimes invoked to explain the second \citep[e.g.][]{Banik2019MNRAS.483.3592B}. The scenario involves a gas cloud forming a single supermassive star instead of many individual stars. This can occur in the presence of local Lyman Werner radiation \citep{Dijkstra2008MNRAS.391.1961D,Agarwal2012MNRAS.425.2854A,Latif2014MNRAS.443.1979L}, baryon dark matter supersonic streaming \citep{Latif2014MNRAS.440.2969L,Schauer2017MNRAS.471.4878S,Hirano2017Sci...357.1375H}, or turbulent cold gas inflows from cosmological scales \citep[][]{Latif2022Natur.607...48L}. The resultant supermassive star (\citealt{umeda2016,woods2017,hammerle2018a}, for a review see \citealt{woods2019}) may be detectable directly \citep{surace2018,surace2019,vikaeus2022}, via a general relativistic instability supernova (GRSN, \citealt{chen2014,whalen2013,nagele2020,moriya2021,Nagele2022arXiv220510493N,Nagele2022arXiv221008662N}), by the observation of gravitational waves \citep{shibata2016,li2018}, or as an ultra long gamma ray burst \citep{sun2017}.

In this paper, however, we consider a different scenario where the supermassive star formation is triggered by a merger of two gas rich galaxies \citep[for a review, see][]{Mayer2019RPPh...82a6901M}. The phenomenon of nuclear gaseous disks forming via multi-scale inflows was first investigated in the context of the M-sigma relation as a means of providing a source of dynamical friction for a SMBH binary in order to assist with its eventual merger \citep[][]{Kazantzidis2005ApJ...623L..67K,Mayer2007Sci...316.1874M}. Since then, it has been shown that not only can this disk influence the behavior of existing SMBHs, but it can also collapse under its own gravity to form a new black hole \citep[][]{Mayer2010Natur.466.1082M}. Confirmation of this scenario may be possible with LISA if the central object maintains sufficient asphericity inherited from the nuclear disk \citep{Zwick2022arXiv220902358Z}. The crucial ingredient as it pertains to the current study is that this scenario is agnostic to the metallicity of the interstellar medium (ISM), meaning supermassive stars will form out of metal enriched gas \citep[][]{Mayer2015ApJ...810...51M}.

The galaxy merger scenario predicts extremely high infall rates (up to $\sim 10^4$ $\msun$/year) over relatively short periods of time ($\sim 10^4$ years) on parsec scales. These rapid inflows then translate to possibly comparable accretion rates \citep[see][]{Mayer2015ApJ...810...51M} onto a supermassive protostar. In the most extreme cases, specifically when both merging galaxies have virial masses greater than $10^{11}$ $\msun$, the protostar will continue to accrete gas until it collapses due to the GR radial instability. However, in the mergers of less massive galaxies, infall rates of order $10^3$ $\msun$/year are expected. The protostars formed by these inflows eventually transition to supermassive stars (M $\sim 10^{3-5}$ $\msun$) due to reservoir exhaustion \citep[][]{schleicher2013,Sakurai2015MNRAS.452..755S,Mayer2015ApJ...810...51M}. In this paper, we focus on the evolution and explosions of these supermassive stars.

The behavior of metal enriched supermassive protostars has been investigated previously \citep{fuller1986,montero2012}. In particular, it was shown that if the protostars collapse due to the GR radial instability, then this collapse can cause an explosion powered by the CNO cycle (which we will term a proton rich or pr-GRSN, to differentiate it from an $\alpha$ process driven GRSN). pr-GRSNe were first investigated in \citet{fuller1986}. They used a 1D post Newtonian (PN) code with a 10 isotope nuclear reaction network and found several exploding models spanning the mass range $5\times 10^5 - 10^6$ $\msun$. The metallicity floor for the lowest mass model was $5\times 10^{-3}$. Subsequently, \citet{montero2012} used a 2D BSSN code with parameterized heating rates to investigate models with similar mass, and they were able to include the effects of rotation. For their non rotating models, they found explosions with the same masses as \citet{fuller1986}, but with slightly higher metallicity.

We extend these studies firstly to take stellar evolution into account, so that we can investigate the eventual collapse or explosion of models which were found to be stable by previous works, and secondly to incorporate more accurate energy generation. We follow the collapse of SMSs with a 1D GR hydrodynamics code coupled to a 153 isotope network. The large network allows us to follow the dynamics of the explosion at higher temperatures, and we thus find a lower metallicity floor than in previous works. After running our simulations, we post process the hydrodynamical trajectories with a 514 isotope network designed to fully follow the low temperature rp-process on the proton rich side. Contrary to the conclusions of previous works, we find that the rp-process can play a critical role in the explosion. 

In Sec. \ref{methods} we outline our numerical procedures for stellar evolution, hydrodynamics, and post processing. In Sec. \ref{results_HOSHI}, we present the results of the stellar evolution simulations. In Sec. \ref{results_HYDnuc}, we present the results of our hydrodynamical simulations, while Sec. \ref{results_post} reports the results of the nucleosynthetic post processing. Sec. \ref{results_obs} details  comparisons to current observations and prospects for future observations. Finally, we conclude in Sec. \ref{discussion}.

\section{Methods}
\label{methods}

In this section, we first describe our initial models and stellar evolution code, then provide details of the GR hydrodynamical code and post processing.

\subsection{Stellar evolution}
\label{HOSHI}

\begin{figure*}
    \centering
    \includegraphics[width=2\columnwidth]{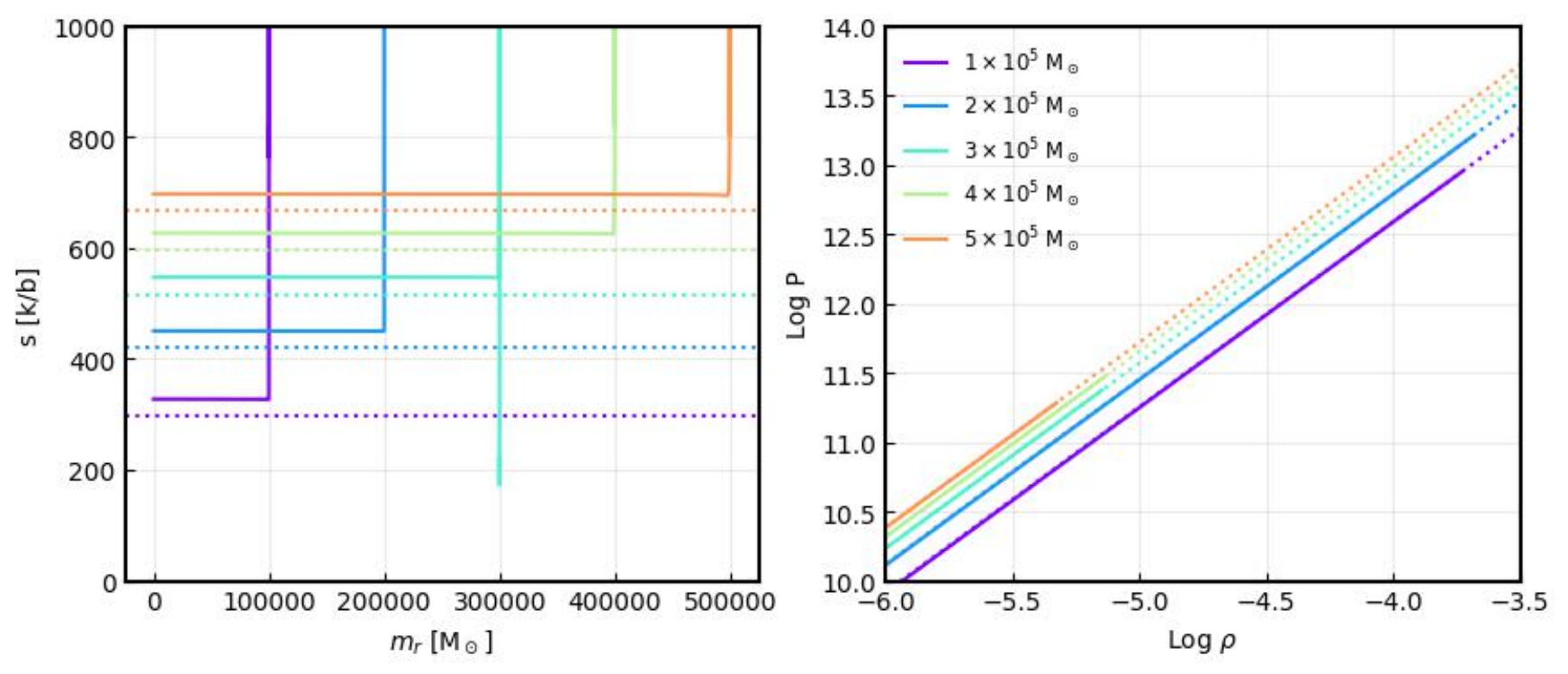}
    \caption{Initial conditions of the HOSHI models for several masses. The left panel shows entropy (solid lines) and radiation entropy (dotted). The right panel shows P-$\rho$ profiles (solid) compared with those of $n=3$ polytropes (dotted).}
    \label{fig:intial}
\end{figure*}

The HOSHI code \citep{takahashi2016,takahashi2018,takahashi2019,yoshida2019} is a 1D stellar evolution code which solves the stellar structure and hydrodynamical equations using a Henyey type implicit method. \citet{nagele2020} introduced the first order PN correction to the hydrostatic terms. The PN approximation is extremely accurate for SMSs in hydrostatic equilibrium because the effects of GR are minor. These minor effects must be included, however, because SMSs are radiation dominated and therefore close to instability. Once the evolution of the star becomes dynamical, HOSHI's lack of a shock capture scheme and the PN dynamical corrections necessitate the use of another code. HOSHI includes a nuclear reaction network (52 isotopes), neutrino cooling, mass loss, and rotation. Mass loss is extremely large and highly uncertain in this regime. We adopt the prescription used in \citet{yoshida2019} and we will show that for most masses and metallicities, mass loss does not overly effect the explosion. The equation of state includes contributions from photons, averaged nuclei, electrons, and positrons. HOSHI uses the Rosseland mean opacity of the OPAL project \citep{Iglesias1996ApJ...464..943I} and solves the Saha equation to determine the ionization of hydrogen, helium, carbon, nitrogen, and oxygen.

In this paper, $M$ is the total mass, $R$ the radius, $T$ the temperature, and $\rho_b$ the baryonic density where quantities with $c$ subscripts show the central values. $s$ is the entropy $s_r$ is the entropy due to radiation at a given mass \citep[][]{shapiro1983}
\begin{equation}
    s_r = 0.942 \bigg( \frac{M}{\msun}\bigg)^{1/2}.
\end{equation}
Finally, X is the mass fraction of a specified element. 

To assist with the analysis, we define various global energy quantities. The internal energy is
\begin{equation}
    E_{\rm int}=\int_0^M \epsilon \; dm_r,
\end{equation}
where $m_r$ is the mass coordinate and $\epsilon$ is the specific energy. The gravitational energy is 
\begin{equation}
    E_{\rm grav}=- \int_0^M g_{\rm effective} \; r \; dm_r,
\end{equation}
where $g_{\rm effective}$ is the local gravity with the 1st order PN correction to the static terms \citep[][]{Nagele2022arXiv220510493N}. The accuracy of this approximation degrades with increasing density and velocity, neither of which are particularly concerning for our purposes. The kinetic energy is 
\begin{equation}
    E_{\rm kin}=\int_0^M \frac{v^2}{2} \; dm_r,
\end{equation}
where $v$ is the radial velocity. The binding energy of the star is the negative of the thermal and gravitational energies (so that a more tightly bound star has higher $E_{\rm bind}$), while the total energy additionally includes kinetic energy:
\begin{equation}
    E_{\rm bind}=-(E_{\rm int}+E_{\rm grav})
\end{equation}
\begin{equation}
    E_{\rm tot}=E_{\rm int}+E_{\rm grav}+ E_{\rm kin}.
\end{equation}
As in our previous works, we define the explosion energy as the total energy at shock breakout. For HYDnuc, we also report the integration over energy generation due to the nuclear network and neutrino cooling (dots indicate time derivatives):
\begin{equation}
    E_{\rm nuc} (t)= \int_{0}^t \int_0^M \dot{\epsilon_{\rm nuc}} \; dm_r \; dt
\end{equation}
\begin{equation}
    E_{\rm \nu} (t)= \int_{0}^t \int_0^M \dot{\epsilon_{\rm \nu}} \; dm_r \; dt.
\end{equation}

We initiate the HOSHI code with a structure resembling an $n=3$ polytrope and having a low central temperature ($T_c < 10^7$ K) and high entropy (that is, larger than than the value for a purely radiative star \citealt{shapiro1983}) as can be seen in Fig. \ref{fig:intial}. Entropies above the purely radiative value are expected for thermally supported supermassive protostars \citep[][]{hosokawa2012}. This low temperature, low density, polytrope then contracts towards the onset of nuclear burning. Because of the presence of carbon, the CNO cycle can stabilize the star without requiring $3\alpha$ reactions around Log $T_c = 8.2$ and the star instead stabilizes around Log $T_c = 7.7$ (this is in contrast to the Pop III case, see e.g. \citealt{woods2020,Nagele2022arXiv220510493N}). Once nuclear burning begins, the entropy decreases to match the radiative value.

\begin{table*}
	\centering
	\caption{Summary of the state of the stellar evolution simulations at the GR instability. The columns are, initial mass, metallicity, and mass loss, followed by several quantities at the GR instability: mass, radius, central temperature, central density, central entropy and its ratio to the radiative value, central hydrogen mass fraction, and binding energy.}
	\label{tab:HOSHI}
	\begin{tabular}{|ccc|r|r|r|r|r|r|r|r|r|r|} 
		\hline\hline
    		M [$10^5$ $\msun$] & Z$/\zsun$ & $\dot{\rm M}$ &M$_f$ [$10^5$ $\msun$] & R [$10^{14}$ cm] & $T_c$ [$10^{7}$ K]&$\rho_c$ [$10^{-3}$ g/cm$^3$] & $s_c$ [kb/b] &$s_c/s_r$ &  X($^1$H) & $E_{\rm bind}$ [$10^{54}$ ergs]    \\
    		\hline
0.5 & 1 & yes  & 0.3468 & 1.387 & 6.577 & 202.2 & 186.4 & 1.063 & 0.1262 & 0.2759\\
0.5 & $10^{-1}$ & yes  & 0.4223 & 1.602 & 13.76 & 1789 & 188.5 & 0.9735 & 2.277e-6 & 0.5161\\
0.5 & $10^{-2}$ & yes  & 0.4923 & 10.85 & 12.21 & 1220 & 192.9 & 0.9232 & 0.001927 & 0.4976\\
1.0 & 1 & no  & 1 & 1.453 & 6.502 & 119.2 & 297.7 & 0.9993 & 0.2005 & 0.7939\\
1.0 & $10^{-1}$ & no  & 1 & 1.012 & 7.387 & 173.6 & 301.8 & 1.013 & 0.3003 & 1.032\\
1.0 & $10^{-1}$ & yes  & 0.9718 & 2.856 & 7.186 & 158.7 & 306.8 & 1.045 & 0.4186 & 1.154\\
1.0 & $10^{-2}$ & yes  & 0.9963 & 0.8019 & 8.481 & 259.4 & 307.7 & 1.035 & 0.4009 & 1.362\\
1.0 & $10^{-3}$ & no  & 1 & 0.6387 & 9.925 & 411.7 & 311.2 & 1.045 & 0.4441 & 1.693\\
1.0 & $10^{-3}$ & yes  & 0.9997 & 0.5608 & 9.812 & 395.4 & 314.6 & 1.056 & 0.5148 & 1.796\\
1.0 & $2\times 10^{-2}$ & yes  & 0.9923 & 2.049 & 8.212 & 238.9 & 301.9 & 1.017 & 0.3222 & 1.185\\
1.0 & $4\times 10^{-2}$ & yes  & 0.9841 & 2.075 & 7.744 & 199.2 & 304.7 & 1.031 & 0.3688 & 1.165\\
1.0 & $6\times 10^{-2}$ & yes  & 0.9529 & 2.628 & 7.737 & 204.2 & 294.3 & 1.012 & 0.2566 & 0.986\\
1.0 & $8\times 10^{-2}$ & yes  & 0.9408 & 3.017 & 7.481 & 184.1 & 295.6 & 1.023 & 0.2764 & 0.9706\\
1.1 & $10^{-1}$ & yes  & 1.036 & 2.598 & 7.398 & 169 & 311 & 1.025 & 0.3012 & 1.071\\
1.2 & $10^{-1}$ & yes  & 1.169 & 1.276 & 7.236 & 147.3 & 335.6 & 1.042 & 0.4376 & 1.412\\
1.3 & $10^{-1}$ & yes  & 1.277 & 0.8369 & 7.153 & 135.1 & 354.8 & 1.054 & 0.5361 & 1.699\\
1.5 & 1 & no  & 1.5 & 1.101 & 6.279 & 84.66 & 378.4 & 1.037 & 0.4213 & 1.538\\
1.5 & 1 & yes  & 1.077 & 0.5428 & 6.565 & 112.8 & 322.3 & 1.042 & 0.2021 & 0.908\\
1.5 & $10^{-1}$ & yes  & 1.467 & 0.9309 & 7.226 & 130.1 & 377.2 & 1.045 & 0.498 & 1.877\\
2.0 & 1 & no  & 2 & 3.016 & 1.917 & 2.06 & 448.6 & 1.065 & 0.7154 & 0.8276\\
2.0 & $10^{-1}$ & yes  & 2 & 0.9452 & 2.97 & 7.659 & 448.6 & 1.065 & 0.7555 & 1.33\\
2.5 & 1 & no  & 2.5 & 4.451 & 1.352 & 0.6466 & 499 & 1.059 & 0.7154 & 0.7215\\
2.5 & 1 & yes  & 2.496 & 3.733 & 0.9993 & 0.2611 & 499.3 & 1.061 & 0.7154 & 0.5351\\
2.5 & $10^{-1}$ & yes  & 2.499 & 3.606 & 0.9845 & 0.2497 & 500.4 & 1.063 & 0.7555 & 0.5463\\
2.5 & $10^{-2}$ & yes  & 2.5 & 3.18 & 1.05 & 0.3023 & 500.7 & 1.063 & 0.7595 & 0.583\\
2.5 & $10^{-3}$ & yes  & 2.5 & 3.169 & 1.046 & 0.2989 & 500.7 & 1.063 & 0.7599 & 0.5811\\
3.0 & 1 & yes  & 2.984 & 4.526 & 0.9727 & 0.2201 & 543.2 & 1.056 & 0.7154 & 0.6163\\
3.1 & 1 & yes  & 3.085 & 4.601 & 0.9764 & 0.2189 & 551.8 & 1.055 & 0.7154 & 0.636\\
3.2 & 1 & yes  & 3.184 & 4.226 & 0.9883 & 0.2234 & 560.1 & 1.054 & 0.7154 & 0.6641\\
3.3 & 1 & yes  & 3.284 & 4.704 & 0.9921 & 0.2225 & 568.3 & 1.053 & 0.7154 & 0.685\\
3.4 & 1 & yes  & 3.385 & 4.563 & 0.9932 & 0.2198 & 576.5 & 1.052 & 0.7154 & 0.7055\\
3.5 & 1 & yes  & 3.491 & 4.572 & 1.112 & 0.3039 & 584.8 & 1.051 & 0.7154 & 0.8112\\
4.0 & 1 & yes  & 4 & 10.36 & 0.9986 & 0.2054 & 624.1 & 1.048 & 0.7154 & 0.8313\\
5.0 & 1 & yes  & 5 & 12.13 & 0.9695 & 0.168 & 694.2 & 1.042 & 0.7154 & 0.994\\
		\hline\hline
	\end{tabular}
\end{table*}

The setup described above is designed to deliver protostars to the main sequence. However, in the context of usual supermassive stars, this represents an approximation of sorts. In reality, supermassive protostars can ignite hydrogen burning before accretion has terminated (e.g. \citealt{hosokawa2013}), a process which we do not model here because HOSHI cannot currently handle accretion. Thus the chemical composition of the star and the structure will have begun to evolve by that point. However, in the galaxy merger scenario, the accretion rates are thought to be extremely large. The fact that the inflows on parsec scales peak at around $10^4 \msun$/yr sets an upper limit for the accretion rate onto the protostar itself. Further work is required to determine the exact accretion rates, but unless they are smaller than the inflows by several orders of magnitude, then hydrogen burning will not have nearly enough time to turn on before the accretion terminates (that is, the relaxation time of the protostar will be longer than the accretion timescale). Note that the accretion timescale in the galaxy merger scenario is quite short compared to other SMS formation scenarios, being order of magnitude $\sim 10^4$ years (see Fig. 3 of \citet{Mayer2010Natur.466.1082M}, Fig. 5 of \citet{Mayer2015ApJ...810...51M}), though again, further work is needed to better constrain this value, especially for the less massive mergers which produce SMSs. Accretion will terminate with the exhaustion of the gas reservoir. Once this termination occurs, the accreting SMS will contract and its structure will resemble the ones considered here. Finally, we note that our models could also be formed by stellar collisions in nuclear star clusters \citep{Denissenkov2014MNRAS.437L..21D}, although the masses are on the upper end of what is thought to be possible via that mechanism.

Once the SMS has settled onto the main sequence, the question of stability must be considered. We determine the stability of the star in HOSHI by solving the pulsation equation for a hydrostatic, spherically symmetric object in general relativity \citep{chandrasekhar1964}:

        \begin{equation}
            \nonumber
            e^{-2a-b}\dv{}{r} \; \Big[ e^{3a+b}\Gamma_1 \frac{P}{r^2} \dv{}{r} \; (e^{-a}r^2\xi)\Big] - \frac{4}{r}\dv{P}{r} \xi +e^{-2a+2b}\omega^2(P+\rho c^2)\xi 
        \end{equation}     
        \begin{equation}
            \label{eq_GR}
            - \frac{8\pi G}{c^4}e^{2b}P(P+\rho c^2)\xi - \frac{1}{P+\rho c^2} \bigg(\dv{P}{r}\bigg)^2 \xi = 0,
        \end{equation}
where $a,b$ are the metric coefficients as defined in \citet{haemmerle2020}, $r$ is the radius, $P$ the pressure, $\Gamma_1$ the local adiabatic index at constant entropy, $\rho = \rho_b(1+\epsilon)$ the relativistic density, and $\epsilon$ the specific internal energy (we absorb rest mass due to mass excess of nuclei into this energy).

The star is unstable if there exists a trial function $\xi(r) \propto e^{i\omega t}$ with $\omega^2 < 0$, representing a perturbation which will grow exponentially. There are two main approaches to solving this equation, either by assuming a nearly linear trial function $\xi \propto r e^a$ \citep{haemmerle2020}, or by iteratively solving for the fundamental mode of the normal mode decomposition of perturbations to Eq. \ref{eq_GR} \citep{Nagele2022arXiv220510493N}. Here we adopt the latter approach, and our exact method is described in Appendix A. For every five timesteps in HOSHI, we solve Eq. \ref{eq_GR} using the profiles from HOSHI. If the result is unstable, we then test that stellar profile in the hydrodynamical code. For the models presented in this paper, all but one of the unstable models either collapse or explode in the hydrodynamical code (Sec. \ref{HYDNUC}). The results of the HOSHI calculations at the point of instability are summarized in Table \ref{tab:HOSHI}.

\subsection{Hydrodynamics}
\label{HYDNUC}

HYDnuc is a 1D Lagrangian GR hydrodynamics code which uses a Roe-type approximate linearized Riemann solver \citep{yamada1997,takahashi2016,nagele2020}. It includes all of the physics from HOSHI except for convection and ionization. In this paper, we use a 153 isotope network (Table \ref{tab:networks}). The 153 isotope network \citep[][]{takahashi2018} covers the proton rich side (p side) at a depth of 3-8 isotopes up to zinc. Reaction rates for all networks are taken from \textit{JINA REACLIB} \citep[][]{Cyburt2010ApJS..189..240C}.

We use the same scheme to transport our models from HOSHI to HYDnuc as in \citet{Nagele2022arXiv220510493N} which is based on the frequency function defined in \citet{takahashi2019}. We do not port the unstable HOSHI models with temperature below $T_c \approx 7$ as these models sometimes run into numerical difficulties in HYDnuc. Due to the exploratory nature of this study and the requirement of a larger nuclear network, we use slightly less optimal numerical parameters than in \citet{Nagele2022arXiv220510493N}, specifically 255 mesh points and $\mathcal{V}=10^{-4}$ (maximum allowed fractional variation of independent variables per timestep). The effect of these changes is to underestimate the energy generated by nuclear burning (see Fig. 6 of \citealt{Nagele2022arXiv220510493N}). As in our previous works, we terminate the simulations when convergence issues arise due to large radius ($r\sim10^{15}$ cm).

\begin{figure*}
    \centering
    \includegraphics[width=2\columnwidth]{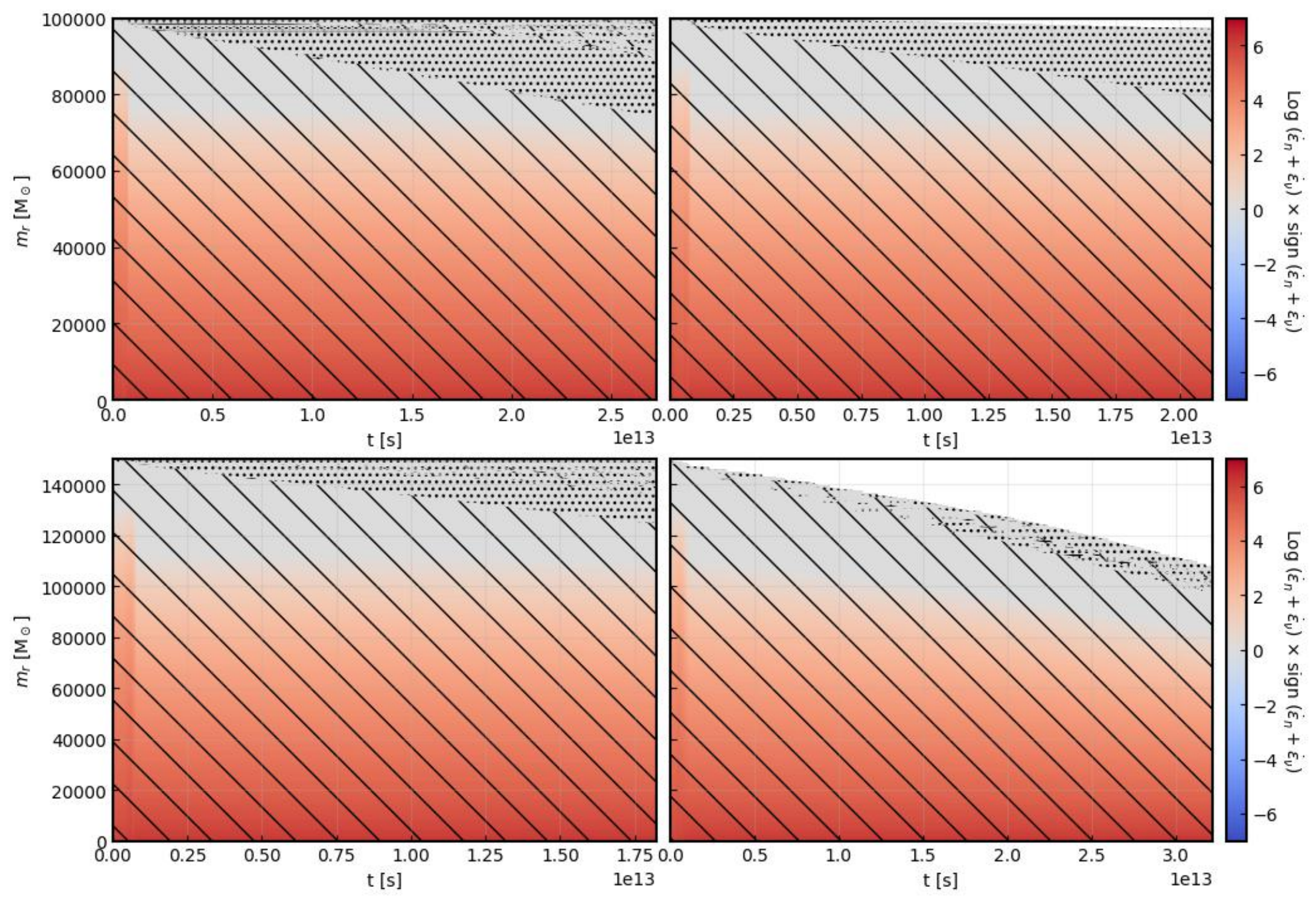}
    \caption{Kippenhahn diagrams with (right) and without (left) mass loss for the $10^5$ $\msun$, $0.1$ $\zsun$ model (upper) and the $1.5 \times 10^5$ $\msun$, $\zsun$ model (lower). Hatches show convective regions (diagonal), radiative regions (dotted) and semi-convective regions (crossed).}
    \label{fig:KH}
\end{figure*}

\begin{table}
	\centering
	\caption{Summary table for the nuclear networks. Entries show the range in A for the specified element.}
	\label{tab:networks}
	\begin{tabular}{|c|l|l|l|c|l|l|l|l} 
		\hline\hline
    		Element &52& 153 & 514 &(cont.) &52& 153 & 514   \\
    		\hline

n&1&1&1&V&47&45-51&42-53\\
H&1-3&1-3&1-3&Cr&48&47-54&44-55\\
He&3-4&3-4&2-4&Mn&51&49-55&46-57\\
Li&6-7&6-7&4-7&Fe&52-56&51-58&48-60\\
Be&7-9&7-9&5-9&Co&55-56&53-59&50-61\\
B&8-11&8-11&6-11&Ni&56&55-62&52-66\\
C&12-13&12-13&8-14&Cu&---&57-63&54-68\\
N&13-15&13-15&10-16&Zn&---&60-64&56-71\\
O&14-18&14-18&12-18&Ga&---&---&58-73\\
F&17-19&17-19&14-20&Ge&---&---&60-75\\
Ne&18-20&18-22&16-22&As&---&---&62-76\\
Na&23&21-23&18-24&Se&---&---&64-81\\
Mg&24&22-26&20-26&Br&---&---&66-82\\
Al&27&25-27&22-28&Kr&---&---&68-86\\
Si&28&26-32&24-30&Rb&---&---&70-87\\
P&31&29-33&26-32&Sr&---&---&72-89\\
S&32&30-36&28-37&Y&---&---&74-91\\
Cl&35&33-37&30-39&Zr&---&---&76-95\\
Ar&36&34-40&32-43&Nb&---&---&78-96\\
K&39&37-41&34-45&Mo&---&---&80-98\\
Ca&40&38-43&36-48&Tc&---&---&82-98\\
Sc&43&41-45&38-49&Ru&---&---&84-99\\
Ti&44&43-48&40-51&\\

		\hline\hline
	\end{tabular}
\end{table}

\begin{table*}
	\centering
	\caption{Summary of the explosions in the hydrodynamical calculation and post processing. The columns are, initial mass, metallicity, and mass loss, followed by maximum central temperature, the change in central hydrogen mass fraction, explosion energy, maximum outflow velocity, and the ejected mass of $^{14}$N, $^{45}$Sc, and $^{56}$Ni in units of $\msun$.}
	\label{tab:explo}
	\begin{tabular}{|ccc|r|r|r|r|r|r|r|r|r|r|} 
		\hline\hline
    		M [$10^5$ $\msun$] & Z$/\zsun$ & $\dot{\rm M}$ & $T_{c,max}$ [$10^{7}$ K]&  $\Delta$X$_c$($^1$H)& $E_{\rm exp}$  [$10^{54}$ ergs] & max(v) [$10^8$ cm/s] & M($^{14}$N)  & M($^{45}$Sc)  & M($^{56}$Ni)      \\
    		\hline
1.0 & 1 & no  & 21.15 & 0.07972 & 7.345 & 8.208 & 331.9 & 0.003939 & 2.687e-15\\
1.0 & $10^{-1}$ & no  & 36.54 & 0.03783 & 8.493 & 10.92 & 80.03 & 0.01034 & 2.501e-7\\
1.0 & $10^{-1}$ & yes  & 33.71 & 0.03388 & 7.971 & 9.764 & 79.53 & 0.001718 & 1.875e-8\\
1.0 & $6\times 10^{-2}$ & yes  & 28.41 & 0.02573 & 4.312 & 8.517 & 64.67 & 0.0004773 & 4.646e-11\\
1.0 & $8\times 10^{-2}$ & yes  & 36.42 & 0.03127 & 7.391 & 10.06 & 545.5 & 0.2503 & 5.965e-6\\
1.1 & $10^{-1}$ & yes  & 41.83 & 0.03798 & 11.41 & 11.63 & 79.76 & 0.2495 & 2.818e-5\\
1.5 & 1 & no  & 27.26 & 0.1574 & 42.03 & 13.25 & 1203 & 0.006465 & 1.14e-11\\
1.5 & 1 & yes  & 27.3 & 0.1213 & 21.68 & 10.12 & 833.4 & 0.004427 & 1.675e-12\\
2.0 & 1 & no  & 27.34 & 0.1778 & 64.62 & 11.41 & 866.4 & 0.009184 & 4.821e-11\\
2.5 & 1 & no  & 30.3 & 0.2007 & 107.1 & 13.57 & 1165 & 0.01735 & 2.732e-9\\
2.5 & 1 & yes  & 23.43 & 0.1775 & 57.52 & 9.177 & 932.9 & 0.009946 & 1.685e-13\\
3.0 & 1 & yes  & 34.01 & 0.2204 & 162 & 15.73 & 1476 & 0.04457 & 2.235e-7\\
3.1 & 1 & yes  & 37.09 & 0.2239 & 186.7 & 16.89 & 1573 & 0.1745 & 5.295e-6\\
3.2 & 1 & yes  & 41.44 & 0.2284 & 213.5 & 18.23 & 1669 & 4.664 & 0.0002803\\
3.3 & 1 & yes  & 47.15 & 0.2359 & 244.6 & 19.67 & 1745 & 18.85 & 0.0222\\
3.4 & 1 & yes  & 59.29 & 0.2511 & 298.3 & 22.07 & 1687 & 25.93 & 180.7\\
		\hline\hline
	\end{tabular}
\end{table*}

\subsection{Post-processing}
\label{POST}

After performing the HYDnuc simulations, we post process the hydrodynamical trajectories using a 514 isotope network (Table \ref{tab:networks}) designed to follow the rp process up to ruthenium. All isotopes on the p side are covered up to the line with slope 1 and intercept 5. Even though this post processing is less computationally expensive, the network is too large to solve the composition at every timestep of HYDnuc. We choose to solve the composition with a frequency of $100^{-1}$ timesteps$^{-1}$, and have checked that a) the convergence of $E_{\rm nuc}$ as the frequency increases and b) that $E_{\rm nuc}$ with frequency $100^{-1}$ agrees with $E_{\rm nuc}$ with frequency $10^{-1}$ to within $0.1\%$. At the end of the HYDnuc simulation, the post-processed composition contains many radioactive isotopes. We then fix the temperature and density and continue to post-process for an additional $10^{12}$ seconds while logarithmically increasing the timestep. $10^{12}$ seconds is enough for most, but not all (e.g. $^{26}$Al) radioactive isotopes to decay.

\section{Results}
\label{results}

In this section, we first describe the results of the HOSHI code, after which we provide details of the hydrodynamics and nucleosynthesis.

\subsection{Stellar Evolution}
\label{results_HOSHI}

Kippenhahn diagrams for typical models, both with and without mass loss, are shown in Fig. \ref{fig:KH}. After nuclear burning arrests the initial contraction and ZAMS is reached, the star is fully convective. In most cases, radiative and semi-convective regions develop inwards from the stellar surface, but in the case of strong mass loss, this process is sporadic  as much of what would have been the radiative region is lost to the stellar wind. All four of the models shown encounter the GR radial instability before the end of hydrogen burning. Mass loss in the hydrogen burning phase is roughly $\sim 0.01 \msun /$yr for the solar metallicity models and lower values for sub-solar metallicity models. In slightly less massive SMSs, mass loss increases after the end of helium  burning and a large fraction of the star is lost \citep[][]{Nagele2023arXiv230405013N}. However, the models in this paper are massive enough that they collapse due to the GR radial instability before this happens.

The models in this paper fall into three broad categories, those that collapse due to the GR instability after hydrogen burning, during hydrogen burning, and before hydrogen burning. Models that collapse after hydrogen burning are either low mass or are mass loss models at solar metallicity. For the solar metallicity models, the $\sim 0.01 \msun /$yr mass loss is large enough to postpone the onset of the GR radial instability. The mass loss models with solar metallicity and M$ \leq 10^5$ $\msun$ do not become GR unstable before the end of the HOSHI calculations. The one model in this study that becomes unstable in between hydrogen and helium burning collapses due to lack of fuel ($5\times 10^4$ $\msun$, $0.1$ $\zsun$) and we did not find any models which become unstable during helium burning.

At sub-solar metallicities, mass loss is weaker and most of the models become unstable during hydrogen burning. However, the exact moment of the onset of the GR radial instability does not follow any obvious trends in mass or metallicity. The structure of the star depends on the complex interplay of energy generation, opacity, and convection all of which depend strongly on metallicity while the last of these also depends stochastically on the initial mass. That being said, all models which collapse during the hydrogen burning phase will explode if they have sufficient metallicity (Sec. \ref{results_HYDnuc}).

The third category is GR unstable from the very beginning of the simulation. These models are massive enough that the GR instability first appears less than two hundred timesteps into the HOSHI simulation and thus their structure at collapse is mostly determined by the initial mass of the model.

\subsection{Hydrodynamics}
\label{results_HYDnuc}

\begin{figure}
    \centering
    \includegraphics[width=1\columnwidth]{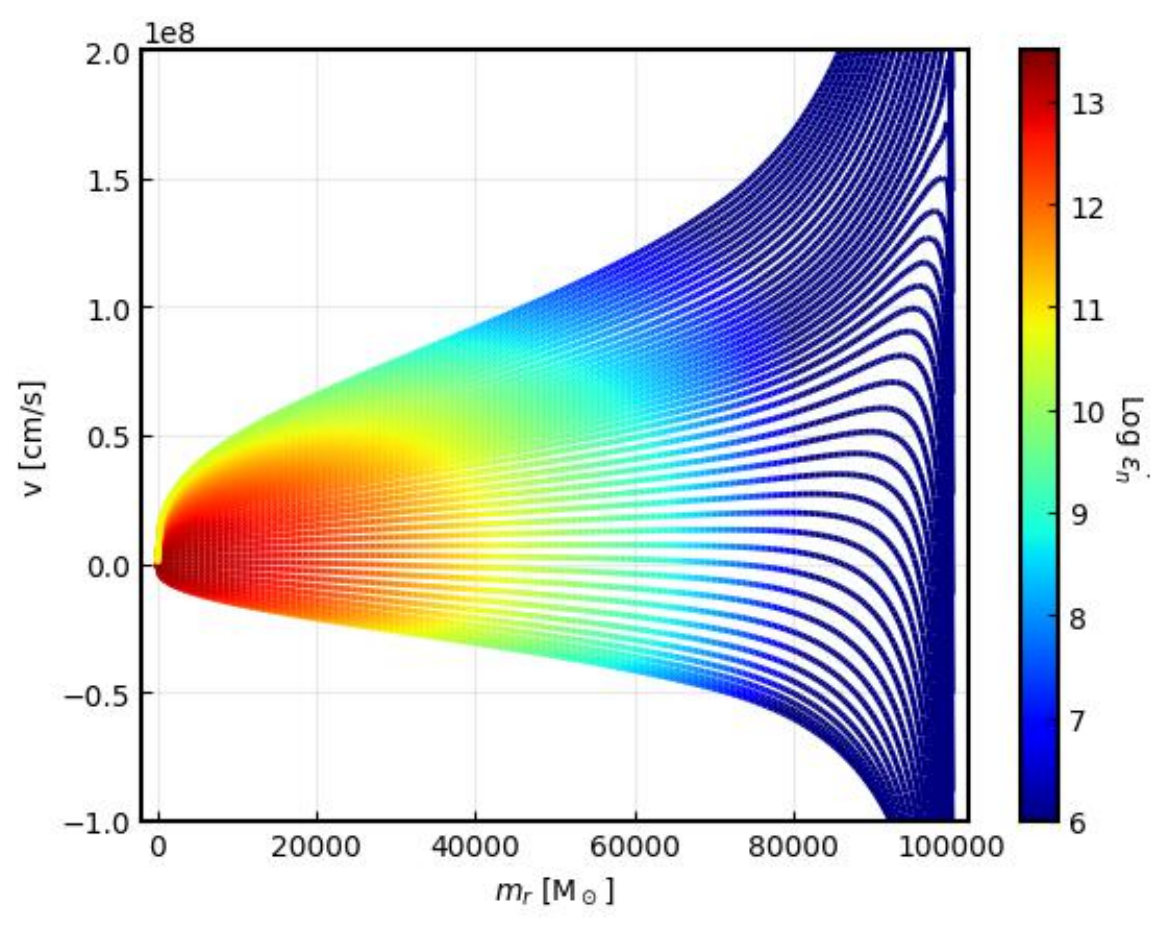}
    \caption{Velocity snapshots for several timesteps near the maximum temperature. Color denotes log heating due to nuclear reactions. }
    \label{fig:rv_fiducial}
\end{figure}

As hinted at in the previous section, the only requirement for an explosion is the existence of sufficient nuclear fuel. If the collapse occurs in the hydrogen burning phase, then this fuel is light elements (particularly CNO) which allow rapid hydrogen burning. If the collapse occurs in the helium burning phase, this fuel is $^{16}$O produced during the helium burning phase as $^{16}$O($\alpha,\gamma$) is the first reaction in the rapid alpha process \citep[][]{Nagele2022arXiv220510493N}. In theory, a SMS with super-solar oxygen might be able to explode via this mechanism without first producing oxygen on evolutionary timescales (that is, if it collapses during the transition to helium burning). Super-solar metallicity is thought to be possible in the galaxy merger scenario \citep[][]{Mayer2023arXiv230402066M}. 

\begin{figure*}
    \centering
    \includegraphics[width=2\columnwidth]{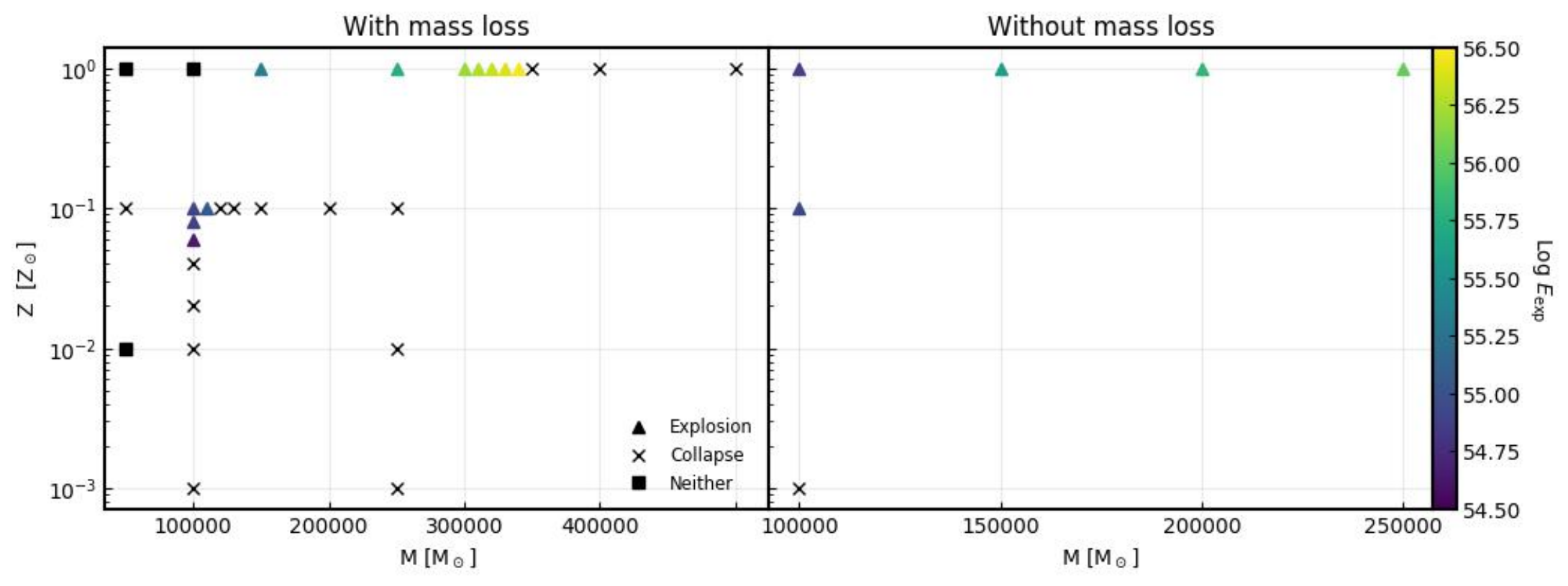}
    \caption{Dependence of explosion energy (color) for the exploding models (triangles) on mass and metallicity. The black crosses are models which failed to explode and black squares either do not reach the GR instability (because of large mass loss) or are stable. The left panel shows models with mass loss while the right panel shows those without.}
    \label{fig:mass_metal}
\end{figure*}

The explosive nuclear burning occurs in a large section of the star, with stronger heating nearer the center (Fig. \ref{fig:rv_fiducial}). As a result of this energy production, the inwards velocity is reversed and the star is eventually unbound. Nuclear reactions continue to occur until the temperature becomes sufficiently low. The explosion energy and maximum outflow velocities of all models are summarized in Table \ref{tab:explo}. Fig. \ref{fig:mass_metal} shows explosion energy (triangles) as a function of initial mass and metallicity for mass loss (left panel) and no mass loss (right panel). The inclusion of mass loss primarily effects the timing of the GR instability, but the explodibility and explosion energy are determined by the mass and metallicity, and thus these largely do not depend on whether or not mass loss is included. As is apparent, more massive models require higher metallicity (more seed metals for proton captures) in order to explode.

\begin{figure*}
    \centering
    \includegraphics[width=2\columnwidth]{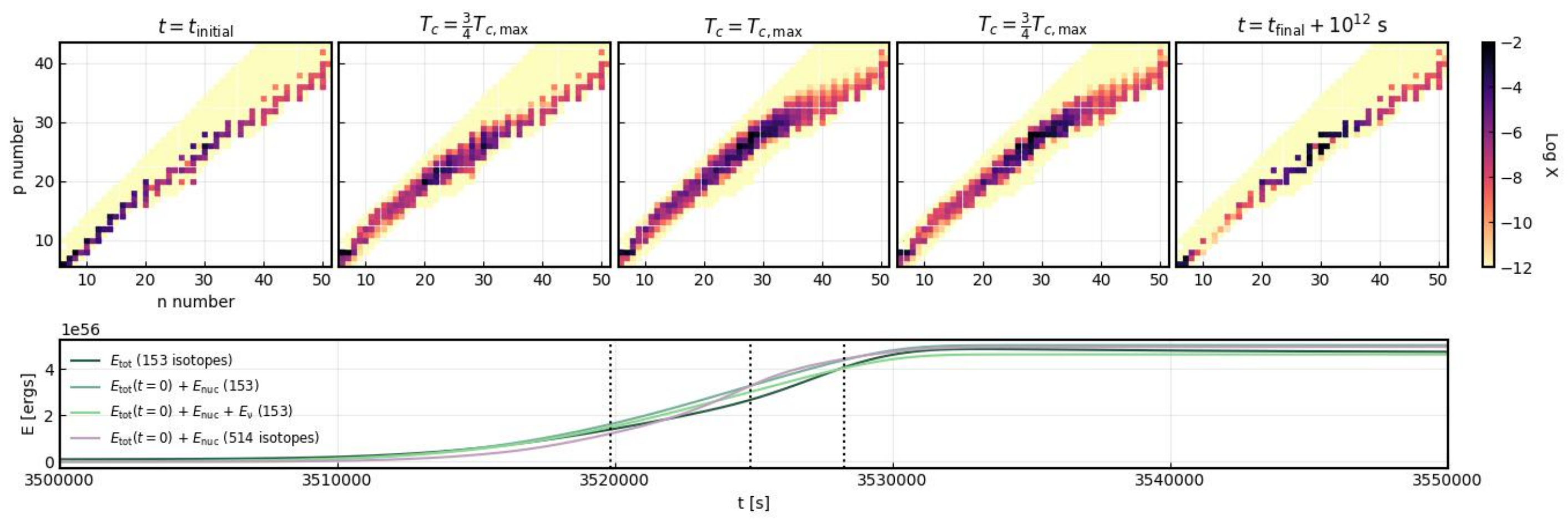}
    \caption{Nucleosynthesis in the post processed 514 isotope network for the massive model, $M=3.4 \times 10^5$ $\msun$, $Z = Z_\odot$. Upper panels --- isotope mass fractions at the central mesh for five snapshots: the initial time, when the temperature rises to 3/4 of the eventual maximum, the maximum temperature, when the temperature falls to 3/4 of the maximum, and $10^{12}$ seconds after the end of the hydrodynamical calculation. Lower panel --- total energy compared to $E_{\rm nuc}$ in both the 153 and 514 isotope calculations and compared to $E_{\rm nuc} + E_\nu$ for the 153 isotope calculation. The vertical dashed lines show the times of the central three panels.}
    \label{fig:rp}
\end{figure*}

In order to estimate a condition for explosion, we compute the amount of energy available due to nuclear reactions in a star with mass M, assuming that the rp process proceeds up to an isotope with proton number m (so we are interested in nuclear potential energy, $P_{\rm nuc, m}$(M)). We assume that every element with atomic number greater than 15 is converted to the most abundant solar isotope with proton number $m$ ($I_m$). Then, the chemical distribution of the star will change as follows:

\begin{equation}
    X_{I_j}' = 
    \begin{cases} 
      X_{I_j} & A(I_j) < 16 \;\rm{or}\;  A(I_j) > A(I_m) \\
      0 &  15 < A(I_j) < A(I_m) 
   \end{cases}
\end{equation}
\begin{equation}
    X_{p}' = X_p - \sum_{I_j} (X_{I_j}-X_{I_j}') \frac{A(I_m) -A(I_j)}{A(I_j)}
\end{equation}
\begin{equation}
    X_{I_m}' = X_{I_m} + X_{p}' + \sum_{I_j} (X_{I_j}-X_{I_j}')
\end{equation}
where unprimed values are the initial mass fractions and primed values are the final ones. $P_{\rm nuc, m}$(M) is the energy released by this change in composition. Then, by comparing this energy to the star's gravitational energy, we estimate that a phenomenological condition for explosion in the $10^5$ $\msun$ models is $P_{\rm nuc, m}(10^5 \;\msun)> 0.15 E_{\rm grav}$. This exercise allows us to test the coverage of the 153 isotope network which has maximum proton number $m=30$. If we had the computational resources to couple the 514 isotope network, this would instead extend to $m=42$. We then check whether any of the collapsing $10^5$ $\msun$ models satisfy the above condition for $m=42$. The Z $=0.04$ $\zsun$ model satisfies the condition, but the Z $=0.02$ $\zsun$ model does not and we hypothesize that extending the nuclear network would not drastically reduce the metallicity threshold for the explosion.

\subsection{Nucleosynthesis}
\label{results_post}

The two main nucleosynthetic processes at work in these explosions are the CNO cycle (lower left 3x3 square of Fig. \ref{fig:rp}) and the low temperature rp process (the rest of Fig. \ref{fig:rp}). The relative importance of these two processes in exploding the star is determined by the maximum temperature (Table. \ref{tab:explo}). The higher the temperature, the more proton captures occur relative to the CNO cycle. In the models in this paper, the higher temperatures associated with the rp process mostly occur in the high mass models ($\geq 3\times 10^5$ $\msun$) or in the central regions of the lower mass, metal poor models. We emphasize that different regions of the star undergo different nuclear reactions (e.g. Fig. \ref{fig:rv_fiducial}).

The yields of these explosions are also characterized by two features. The first is enhanced nitrogen (relative to solar, \citealt{asplund2009}) as well as suppressed carbon and oxygen due to the CNO cycle producing roughly equal amounts of its eponymous elements. The second is a broad exchange of light elements (flourine, sodium, magnesium) for slightly heavier elements (aluminium through vanadium) due to multiple proton captures which then decay back to stability at higher mass number than they originated. The only exception to this trend appears to be neon, which experiences proton captures, but is replenished from below by material exiting the CNO cycle.

\begin{figure}
    \centering
    \includegraphics[width=\columnwidth]{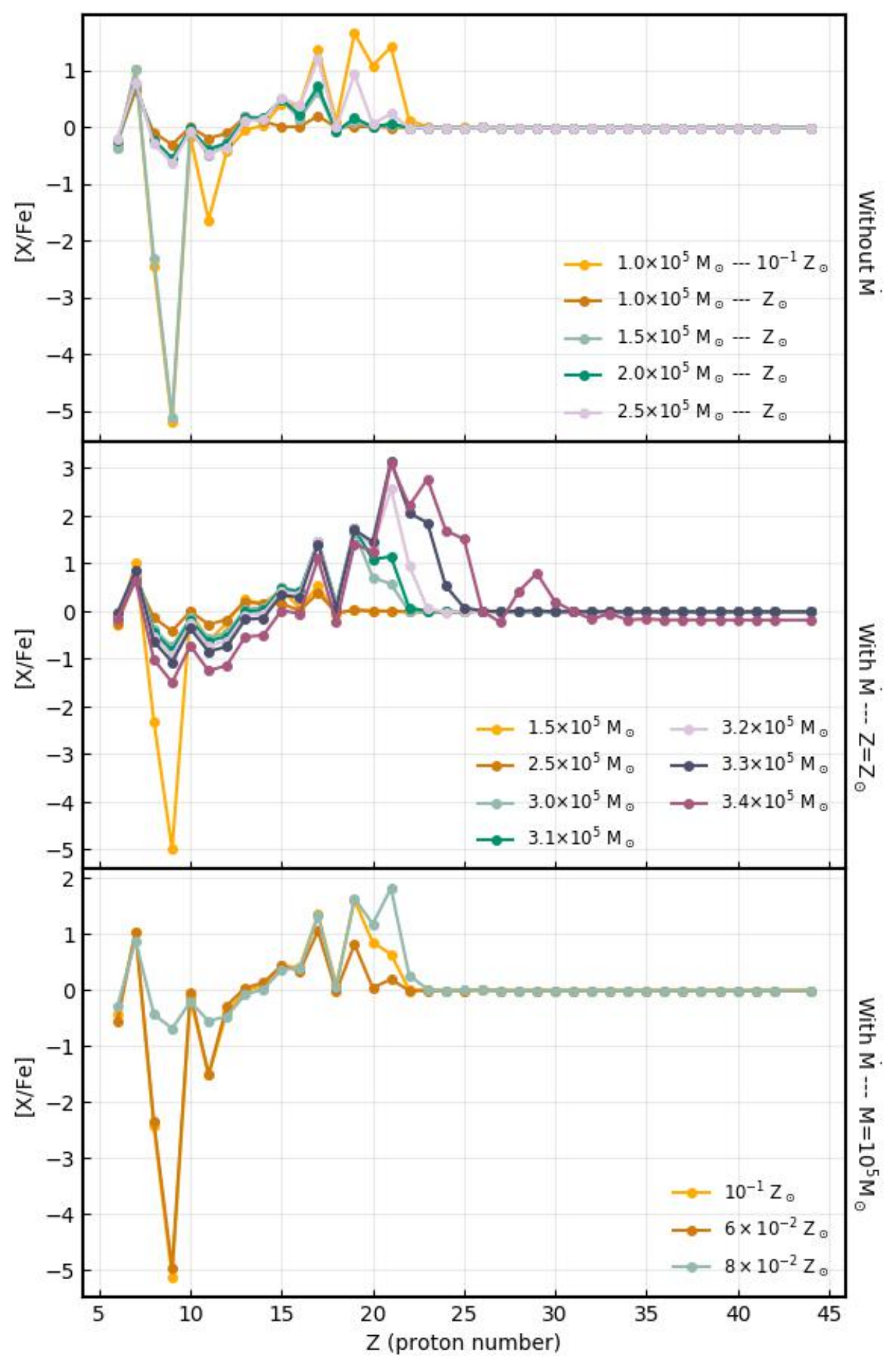}
    \caption{Elemental yields ($t=t_{\rm final}+10^{12}$ s) from the 514 isotope network for models without mass loss (upper panel), solar metallicity models with mass loss (central panel) and $M=10^5$ $\msun$ models with mass loss (lower panel).}
    \label{fig:metal_yields}
\end{figure}

In Fig. \ref{fig:metal_yields}, we show the elemental yields of models without mass loss (upper panel), those with mass loss and solar metallicity (central panel) and those with $10^5$ $\msun$ (lower panel). Besides the fact that the higher mass models reach higher temperatures, there are not apparent differences in these divisions. All models show excess nitrogen and most show extremely sub-solar flourine. Additionally, these explosions could be identified by detections of odd numbered intermediate mass elements, particularly chlorine, potassium, scandium and vanadium. 

\begin{figure*}
    \centering
    \includegraphics[width=2\columnwidth]{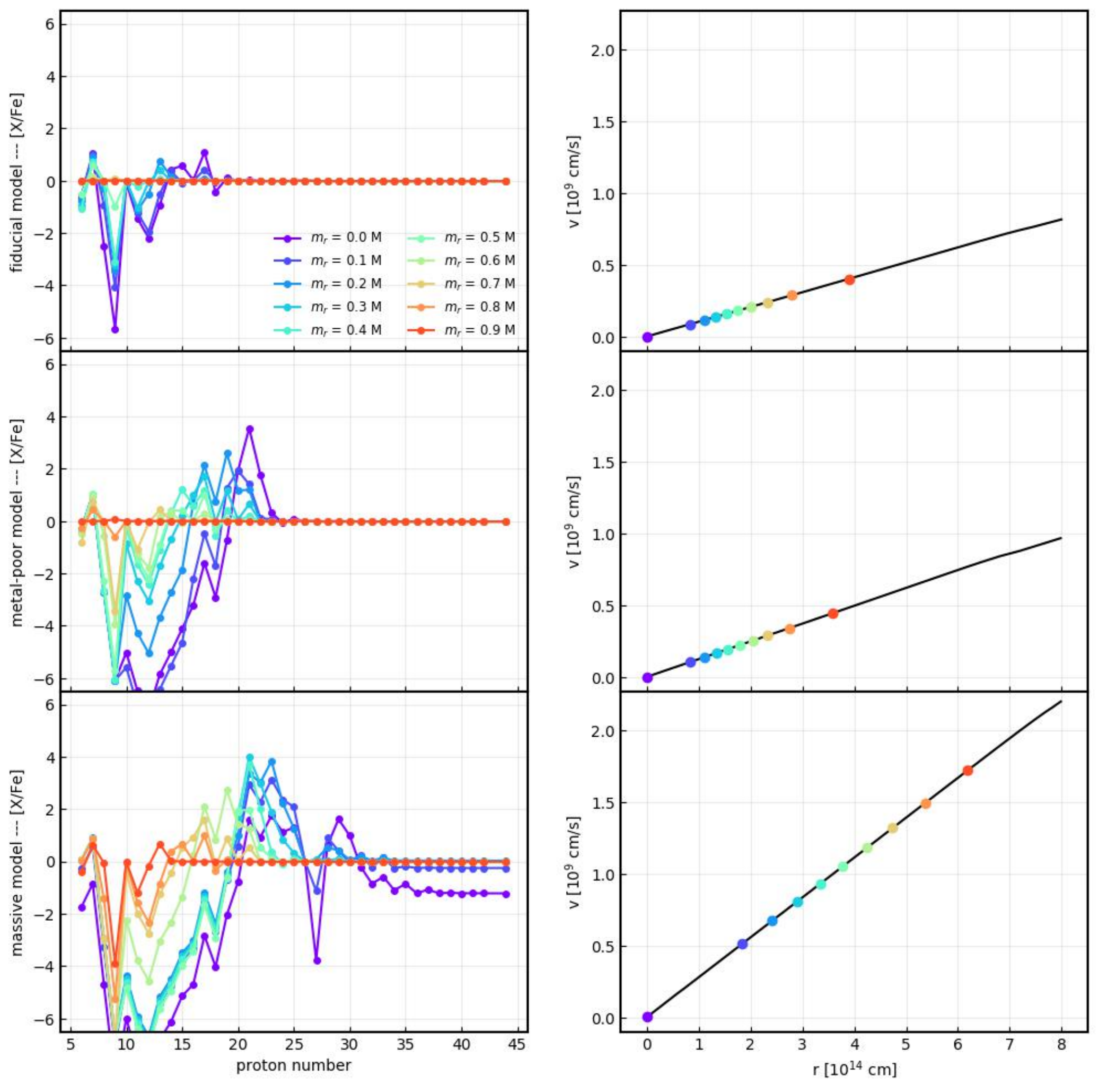}
    \caption{Elemental yields at selected mass coordinates and velocity profiles for the three indicated models. Left column --- elemental abundance for mass coordinates (not the average between mass coordinates) at multiples of $0.1$ times the total mass. Right column --- velocity profiles at a comparable timestep near the end of the HYDnuc simulation for each of the three models. The velocity profiles are nearly homologous. Colored circles indicate the mass coordinate shown in the left column.}
    \label{fig:chemical_feedback}
\end{figure*}

Finally, we consider spatial variation of these yields (Fig. \ref{fig:chemical_feedback}) for a fiducial model (M = $10^5$ $\msun$, Z = $\zsun$, without mass loss), a metal poor model (M = $10^5$ $\msun$, Z = $0.08$ $\zsun$, with mass loss) and a massive model (M = $3.4\times 10^5$ $\msun$, Z = $\zsun$, with mass loss). As expected, the central regions which are expanding less rapidly (nearly homologous expansion) have deficiencies in light elements and super-solar intermediate mass elements. In the fiducial model, nitrogen enhancement is only seen in the inner half of the star, whereas for the other two models, it extends to about $80\%$ of the star. In the central $10\%$ of the massive model, cobalt is suppressed because its lightest stable isotope ($^{59}$Co) cannot be reached from the p side. Also in this region, $^{56}$Fe is synthesized (Table \ref{tab:explo}) which shifts the heavy mass elements to sub-solar abundances (absolute abundances do not change).

\subsection{Rate estimate and prospects for observation}
\label{results_obs}

We now turn to the question of how frequently these pr-GRSN occur. \citet{Bonoli2014MNRAS.437.1576B} calculated the number density of massive galaxy mergers ($M_{\rm halo}>10^{11}\;\msun$) which fulfilled the criteria for merger induced direct collapse (Fig. 4), which is order of magnitude $\phi = 10^{-4}$ [cMpc$^{-3}$ Gyr$^{-1}$] (note that Fig. 4 of \citet{Bonoli2014MNRAS.437.1576B} has units of [cMpc$^{-3}$ 0.1 Gyr$^{-1}$]). As discussed in \citet{Bonoli2014MNRAS.437.1576B}, relaxing the mass asymmetry condition further would increase the merger rate by an order of magnitude. Another increase could be had by relaxing the mass constraint. Indeed, the supermassive stars considered in this paper may be the result of mergers between slightly less massive galaxies, which are thought to occur more often \citep[e.g.][]{OLeary2021MNRAS.501.3215O}. However, it should be noted that the rate assumed in \citet{Bonoli2014MNRAS.437.1576B} would severely overproduce SMBHs and can thus be regarded as an extreme upper limit on the pr-GRSN rate. In a similar vein, a lower limit for this rate can be found by assuming one pr-GRSN per high redshift quasar, which translates to a rate of $ 10^{-9}$ [cMpc$^{-3}$ Gyr$^{-1}$] \citep[e.g.][]{Fan2022arXiv221206907F}.

The most promising avenue for detection of these events is the observation of the supernova itself. We will examine this in more detail in future work, but here we present basic considerations. These explosions are more energetic than the $\alpha$ process GRSN found in other works \citep[][]{chen2014,nagele2020,Nagele2022arXiv220510493N} while also being closer in distance, due to the fact that the SMS is not metal free. Since the $\alpha$ process GRSN can be detected out to redshifts of 10-30 \citep[][]{moriya2021,Nagele2022arXiv221008662N}, it seems likely that supernovae from the explosions in this paper will be easier to detect. 

However, the supernova is by no means the only prospect of identifying one of these explosions. Explosions in this energy range are thought to reach distances of several hundred kpc before falling back into the halo and igniting a violent starburst \citep{Whalen2013ApJ...774...64W,Johnson2013ApJ...775..107J,Whalen2013ApJ...777...99W}. One high redshift, luminous galaxy undergoing a starburst is GN-z11 \citep{Oesch2016ApJ...819..129O,Jiang2021NatAs...5..256J,Tacchella2023arXiv230207234T}. This galaxy was recently observed by NIRCam to have a haze, although a definite association with the galaxy is not confirmed \citep[][]{Tacchella2023arXiv230207234T}. This haze could be evidence of a past merger or a supernova remnant. GN-z11 has also been observed by NIRSpec to have strong nitrogen lines \citep[][]{Bunker2023arXiv230207256B} which have been tentatively interpreted as super-solar nitrogen \citep[][]{Cameron2023arXiv230210142C}. If confirmed, both the haze and the super-solar nitrogen would be circumstantial evidence of a pr-GRSN. Convincing evidence may be hard to nail down, but observations of Cl, K, Sc, or V would be a step in that direction. Other explanations for the super-solar nitrogen include tidal disruption events, winds from evolved stars, peculiar supernovae yields \citep[][]{Cameron2023arXiv230210142C} and winds from SMSs \citep[][]{Charbonnel2023arXiv230307955C,Nagele2023arXiv230405013N}.

In another high redshift observation, \citet{Yoshii2022ApJ...937...61Y} reported a quasar with [Mg/Fe] =-1.11 ± 0.12 at redshift 7.54. They discuss how it is challenging to produce so much iron so early in the Universe. They hypothesize that certain types of pair instability supernovae \citep{takahashi2018} could reproduce this ratio, but the pr-GRSN could also naturally explain this phenomenon because it consumes a large amount of magnesium (Figs. \ref{fig:metal_yields},\ref{fig:chemical_feedback}). At this redshift, the ISM metallicity can be greater than the explosion threshold \citep[][]{Pallottini2014MNRAS.440.2498P}. Closer to home, a large population of nitrogen-enhanced mildly metal poor stars has recently been observed \citep{Fernandez-Trincado2020ApJ...903L..17F}, but these stars do not appear to have other signatures of the pr-GRSN.

\section{Discussion}
\label{discussion}

As the saying goes, hydrogen is flammable. We have shown that if SMSs have sufficient seed metals when they collapse, then they will explode through a combination of the CNO cycle and rp process. The consequences of this are as follows.

Our results present a challenge to the galaxy merger scenario in two senses of the word. The first is that if a SMS can form from the merger of two gas rich galaxies, the mass of this SMS must be sufficiently large in order to collapse to a black hole and form the seed of a high redshift quasar (Sec. \ref{results_HYDnuc}). Furthermore, it is feasible that if the GR radial instability triggers during the accreting phase, a similar explosion might occur, and we plan to investigate this possibility in future work. The second challenge is that galaxy mergers are fairly common events (Sec. \ref{results_obs}). If quasars with masses of $10^9$ $\msun$ are truly being seeded by the most massive mergers, then there should be less massive objects, specifically SMSs, produced by less massive mergers. The question, then becomes, why do we not observe these SMSs, either directly or as pr-GRSNe? In other words, the galaxy merger scenario must posses a way to suppress SMS explosions at low redshift. Such a mechanism is naturally built into other SMS formation scenarios, such as atomic cooling halos, by the stipulation that the gas must be nearly metal free (Z/$\zsun < 10^{-3}$, \citealt{Chon2020MNRAS.494.2851C,Hirano2022arXiv220903574H}). Possibilities for suppressing SMS formation via galaxy mergers at lower redshifts do exist. For instance, if a black hole already inhabits the nuclear disk, then the gas could accrete directly onto the black hole instead of forming a SMS. \citep[][]{Mayer2019RPPh...82a6901M}.

As we have touched upon, the intermediate mass black hole population, or relative lack thereof, also presents a challenge to the study of black holes. If SMBHs originate primarily from the galaxy merger scenario, then our results present a natural explanation for the lack of observed intermediate mass black holes. This is because metal enriched supermassive stars and protostars which are not massive enough to collapse to supermassive black holes (say, $10^6\;\msun$) do not then collapse to intermediate black holes, but instead either explode in a pr-GRSN or lose most of their mass due to line driven winds after the helium burning phase.

We will now summarize some assumptions and shortcoming of the current study. First we will discuss assumptions made in the HOSHI code. We adopt a mass loss prescription which may deviate significantly from the true mass loss rate. Unfortunately, there is little that can be done to address this issue, given the difficulty in observing SMSs directly. In addition, we have only considered non rotating models, but rotation may stabilize the SMSs \citep[][]{Haemmerle2021A&A...650A.204H} and increase the mass loss. Both of these factors would act to prevent the SMS from reaching the GR radial instability. Finally, we do not include the accretion phase in the HOSHI calculations, meaning there is some uncertainty as to how accurate our initial conditions are.

Another shortcoming of this study is that we do not have the resources to couple the 514 isotope network to our GR hydrodynamics code. Thus, all of the nucleosynthesis that we present is the result of post-processing, which may not be entirely accurate. In addition, since the energy generation by the 153 isotope network is smaller than that of 514 (Sec. \ref{methods}), we slightly underestimate the region covered by the pr-GRSN (Fig. \ref{fig:mass_metal}). 

We have shown that supermassive stars which encounter the GR radial instability before the end of hydrogen burning will explode in a pr-GRSN if their metallicity is high enough. The supernovae associated with these events should be visible and these explosions will also leave distinct chemical imprints on their host galaxies. It is likely that current and future surveys with unprecedented breadth and depth will be able to constrain the population of merger induced DCBHs based on the observation or non observation of pr-GRSNe.

\section*{Data Availability}

The data underlying this article will be shared on reasonable request to the corresponding author.

\section*{Acknowledgements}

This study was supported in part by the Grant-in-Aid for the Scientific Research of Japan Society for the Promotion of Science (JSPS, Nos. JP21H01123, JP22K20377).




\bibliographystyle{mnras}
\bibliography{bib}




\section{Appendix A}

This appendix outlays our updated approach for iteratively solving the pulsation equation for a hydrostatic, spherically symmetric fluid in general relativity \citep{chandrasekhar1964,haemmerle2020,Nagele2022arXiv220510493N}:

        \begin{equation}
            \nonumber
            e^{-2a-b}\dv{}{r} \; \Big[ e^{3a+b}\Gamma_1 \frac{P}{r^2} \dv{}{r} \; (e^{-a}r^2\xi)\Big] - \frac{4}{r}\dv{P}{r} \xi +e^{-2a+2b}\omega^2(P+\rho c^2)\xi 
        \end{equation}     
        \begin{equation}
            - \frac{8\pi G}{c^4}e^{2b}P(P+\rho c^2)\xi - \frac{1}{P+\rho c^2} \bigg(\dv{P}{r}\bigg)^2 \xi = 0,
        \end{equation}

There exist a sequence of solutions to this equation $\omega_i$, $\xi_i$ such that the integer $i$ is the number of nodes of $\xi$. If any of these $\xi_i$ correspond to an $\omega_i$ with $\omega_i^2 < 0$, then the perturbation $\xi_i$ will grow exponentially (instead of oscillating) and the star is unstable against this perturbation. Thus, finding any $\omega_i^2 < 0$ is a sufficient condition for instability. This equation is self adjoint and therefore obeys the properties of Sturm-Liouville theory, which implies $\omega_0 < \omega_1 < \omega_2 < ... $. It follows that a star is stable if and only if 

\begin{equation}
    \omega_0^2 < 0.
\end{equation}

In \citet{Nagele2022arXiv220510493N}, we solved the pulsation equation for an initial guess of $\omega_{0,1}^2 = 10^{-7}$. For this value of $\omega_0^2$, we integrate the equation once from the center and once from the surface. We then computed the Wronskian divided by $r$ at a matching radius.
\begin{equation}
     \mathcal{X}(p) = \frac{2\mathcal{W}(p)}{\xi_{\rm in}(p)+\xi_{\rm out}(p)} 
\end{equation}
If the Wronskian is less than a certain threshold (that is, if the two solutions agree), then $\omega_0^2$ is the correct frequency. In practice, it will not be correct, and we repeat this process for $\omega_{0,2}^2 = 0.9 \times 10^{-7}$. Then, we extrapolate a new guess for $\omega_{0}^2$:
\begin{equation}
    \omega_{0}^2 = \frac{\omega_{0,1}^2-\omega_{0,2}^2}{ (\mathcal{X}_2-\mathcal{X}_1)/\mathcal{X}_2}
\end{equation}
The above process is repeated for the new $\omega_0^2$ until the Wronskian of the new solution falls below the threshold. Finally, the corresponding $\xi$ is checked to make sure it is $\xi_0$ (that it has no modes).

This procedure worked for most timesteps of our numerical models (both stellar evolution and hydrodynamical), yet it had several obvious problems. First was the fact that the initial guess for the frequency squared was greater than zero and quite large in absolute terms. This meant that models with $\omega_{0}^2 < 0$ would often take extremely large computational times before converging, and in practice these timesteps were often reported as having found no solution. This also meant that our Wronskian threshold was higher than would be ideal, which caused our accuracy to suffer. Indeed, Fig. 3 of \citet{Nagele2022arXiv220510493N} shows the performance of our approach on numerical polytropes and although the values of $\omega_0^2$ decrease for increasing polytrope resolution, the absolute value (which should be consistent with zero) is $\sim 10^{-8}$. Both of these issues (positive bias, low accuracy) stemmed from our somewhat naive approach to extrapolating the new value of $\omega_0^2$.

\begin{figure}
    \centering
    \includegraphics[width=\columnwidth]{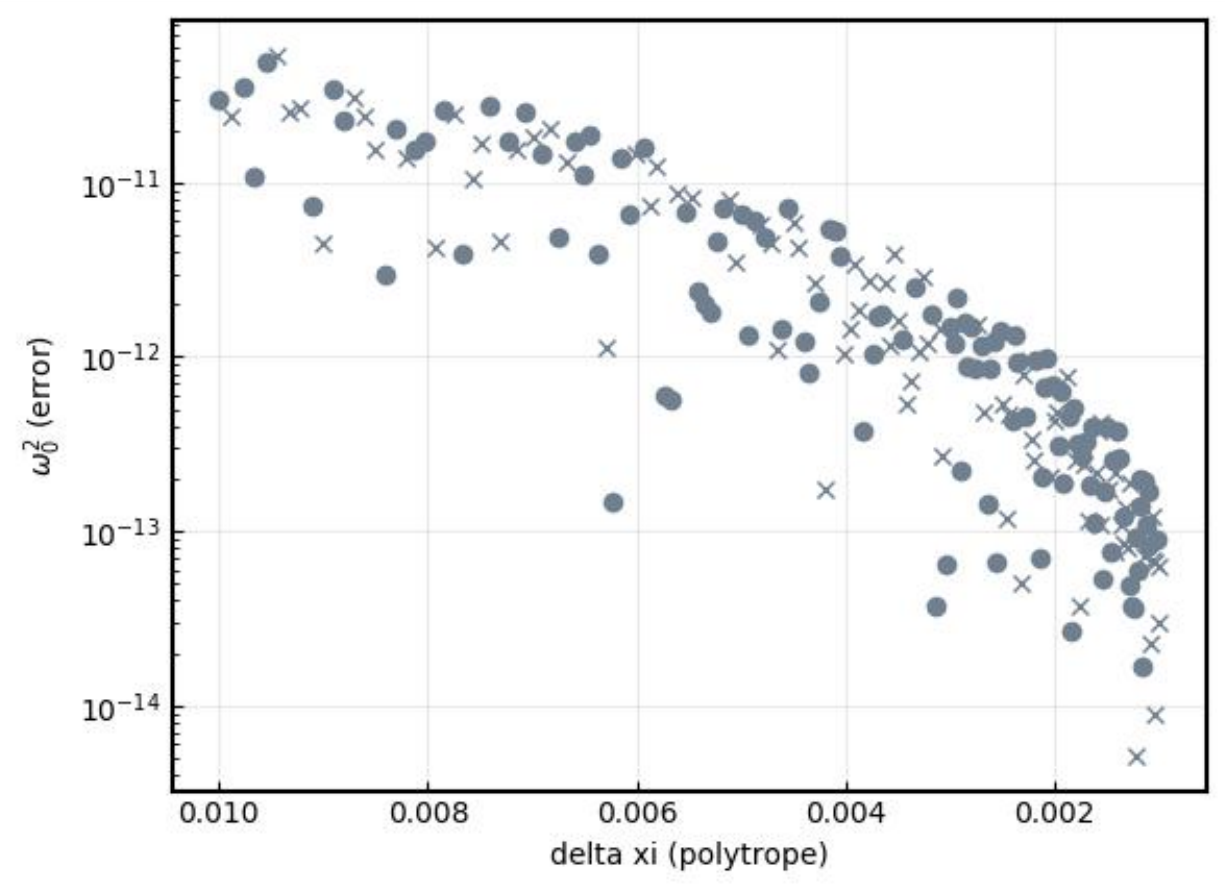}
    \caption{Convergence of the GR stability criterion for numerical $n=3$ polytropes. Crosses are positive values of $\omega_0^2$ and circles negative ones.}
    \label{fig:append1}
\end{figure}

In this paper, we adopt a scheduled extrapolation, so that the above equation becomes:
\begin{equation}
    \omega_{0}^2 = \frac{\omega_{0,1}^2-\omega_{0,2}^2}{ (\mathcal{X}_2-\mathcal{X}_1)/\mathcal{X}_2} \times 
    \begin{cases}
       10^{-2} &\quad   \mathcal{X}_2 / \mathcal{X}_{\rm initial} > 10^{-4} \\ 
       10^{-1} &\quad 10^{-4} > \mathcal{X}_2 / \mathcal{X}_{\rm initial} > 10^{-2} \\        
       1 &\quad \mathcal{X}_2 / \mathcal{X}_{\rm initial} < 10^{-2} \\ 
     \end{cases}.
\end{equation}
$\mathcal{X}_2$ decreases during the iteration process, so the above cases are satisfied from top to bottom. This allows us to set our initial guess to be very small ($\omega_{0}^2 = 10^{-15}$) because the first iteration has a much lower jump than the full extrapolation, and so there is no danger of overshooting the true value. These two changes dramatically increase the accuracy of $\omega_0^2$ for marginally unstable numerical polytropes, to the level of $\sim 10^{-13}$ (Fig. \ref{fig:append1}). Note that at least some of this improvement is due to our recalculation of the parameter $\kappa$. \citet{chandrasekhar1964} had determined this to be $\kappa = 1.1245$ for $n=3$ polytropes, whereas we compute it to be $\kappa = 1.1276$.

This five order of magnitude increase in accuracy for the numerical polytropes does not translate directly to numerical models obtained from simulations because of the finite resolution of the simulations. The results of unstable configurations in both the stellar evolution simulations and the hydrodynamical simulations now appear to oscillate between stability and instability, though why exactly these oscillations occur remains a mystery. 

It does, however, highlight a weakness of this stability analysis, which is meant to be applied at a single moment in time. A numerical model of a star may thus be unstable, but as the star contracts, the structure changes and the star may end up in a configuration which is stable. If at this point the inwards velocity is not too large, then the star may stabilize. Thus, we must use the instability condition in tandem with a relativistic hydrodynamics code in order to know the final fate of an unstable supermassive star.

\bsp	
\label{lastpage}
\end{document}